\documentclass[
  aps,
  prd,
  reprint,
  raggedbottom,
  amsmath,
  amssymb,
  nofootinbib,
  longbibliography
]{revtex4-2}

\usepackage[T1]{fontenc}
\usepackage{hyperref}
\hypersetup{colorlinks=true,linkcolor=blue,citecolor=blue,urlcolor=blue}
\newcommand{\dd}{\mathrm{d}}
\newcommand{\AdS}{\mathrm{AdS}}
\newcommand{\NH}{\mathrm{NH}}

\newcommand{\cA}{\mathcal{A}}
\newcommand{\cH}{\mathcal{H}}
\newcommand{\eps}{\varepsilon}

\begin{document}
\title{Extremal black hole throats from non extremal near horizon expansions}
\author{Anirudhda Shinde}
\email{24dr0031@iitism.ac.in}
\author{Mangesh Mandlik}
\email{mandlik@iitism.ac.in}
\affiliation{Department of Physics, Indian Institute of Technology
(Indian School of Mines), Dhanbad, Jharkhand, India}
\begin{abstract}
We show how extremal near horizon geometries can be recovered from the non extremal near horizon expansion for smooth stationary black hole families admitting a smooth extremal limit. In Gaussian null coordinates common to the family, we apply the near extremal scaling of the surface gravity together with a radial scaling and a rescaling of Killing time. Although the physical surface gravity vanishes, the scaled horizon can retain a finite surface gravity. The extremal near horizon geometry is obtained when this scaled surface gravity is taken to zero. Under the near extremal scaling and the time rescaling, the quadratic radial term in the temporal metric component can survive, while higher radial orders vanish. Expansion through second order is therefore sufficient and is necessary whenever the quadratic term has a nonzero limiting coefficient in the chosen chart. We derive a common pullback prescription for extracting the near horizon coefficients and recover the known throats of Reissner--Nordström, Kerr, Kerr--Newman, and charged rotating Bañados--Teitelboim--Zanelli (BTZ) black holes. Neutral rotating BTZ provides a case in which the string Carroll expansion already suffices. 
\end{abstract}
\maketitle
\section{Introduction}
\label{sec:introduction}

Near horizon limits isolate geometric structures shared by black holes
with otherwise different global properties. For an extremal black hole,
the vanishing surface gravity permits a simultaneous rescaling of radial
distance and Killing time that produces a nondegenerate Lorentzian
near horizon metric \cite{Reall:2002bh,KunduriLuciettiReall,KunduriLucietti}. Familiar examples are the
$\AdS_2\times S^2$ throat of extremal Reissner--Nordstr\"om (RN),
locally the Bertotti--Robinson geometry~\cite{Bertotti1959,Robinson1959,SpradlinStrominger}, and
the near horizon extremal Kerr (NHEK) geometry~\cite{BardeenHorowitz,Guica:2008mu}.
At fixed nonzero surface gravity, the leading longitudinal geometry is
instead Rindler-like~\cite{BagchiII}, and the extremal near horizon
rescaling gives a divergent temporal component in Gaussian null
coordinates~\cite{FontanellaNonextremal}. Relating these descriptions requires control over
both the distance from the horizon and the approach to extremality~\cite{CarrollJohnsonRandall,BengtssonHolstJakobsson}.

Carroll geometry describes the degenerate intrinsic geometry of null
hypersurfaces~\cite{DuvalCarroll,CiambelliNullGeometry}, including black hole
horizons~\cite{DonnayMarteau}. The string Carroll expansion describes the near horizon region of a non extremal black hole through first radial order~\cite{BagchiI,BagchiII}. Under a radial near horizon scaling, the transverse horizon metric remains at leading order, while the longitudinal sector, comprising the time and radial directions, appears at the first subleading order. The metric at leading order is therefore degenerate in these two directions. This provides a geometric framework for organizing the expansion
and for studying strings near non extremal horizons. String probes have been studied in Schwarzschild~\cite{BagchiI,BagchiII}
and BTZ backgrounds~\cite{BanerjeeBTZCarroll}; the RN and Kerr background
expansions are also discussed in ~\cite{BagchiII}.
Particle and scalar probes in more general black hole geometries have
been studied in ~\cite{BagchiLookingGlass}.
In particular, a Gaussian null formulation makes the relation between
the string Carroll expansion and local non extremal horizon data
explicit~\cite{BagchiLookingGlass}.
For extremal horizons, radial expansions in Gaussian null coordinates have also been used to study transverse deformations of the near horizon geometry \cite{LiLuciettiTransverse}.

The approach to extremality raises a further question: how can the Lorentzian near horizon geometry of an extremal black hole be recovered from the non extremal string Carroll expansion? The need to understand extremal limits within the string Carroll description was
identified in ~\cite{BagchiII,BanerjeeBTZCarroll,BagchiLookingGlass}.
Simply sending the surface gravity to zero within this first order expansion does not, in general, recover the extremal throat. In this paper we consider a limit in which the approach to the outer horizon and the approach to extremality occur simultaneously. We implement this by scaling the surface gravity with the same parameter that controls the radial expansion, and refer to this as the near extremal scaling. Together with a rescaling of Killing time, this reorganizes the radial expansion: terms suppressed at fixed black hole parameters can contribute at leading order in the limiting geometry. 

For RN, this issue was studied in ~\cite{ShindeMandlik}. In ingoing
Eddington--Finkelstein coordinates, the string Carroll expansion does not
retain all the data needed for the throat: a quadratic contribution to
the temporal component survives the near extremal scaling and the time rescaling, and supplies
the missing curvature. Retaining second order recovers the scaled
throat, and sending its surface gravity to zero gives
$\AdS_2\times S^2$. In static coordinates, recovering the radial
component requires contributions from arbitrarily high radial orders.
This coordinate dependence motivates
a formulation in which regularity at the horizon and control of the
radial expansion are explicit.

 We extend the RN analysis to smooth stationary black hole families admitting a smooth extremal limit. We work in a common Gaussian null chart in which the metric coefficients remain smooth through extremality. We show that the near horizon expansion through second radial order suffices to recover the limiting metric and give a pullback prescription for extracting the required coefficients from horizon regular seed metrics. The extremal throat is obtained
when the surface gravity in the rescaled time normalization vanishes.

We formulate the construction for generic stationary black holes in arbitrary spacetime dimension and illustrate it with charged and rotating black holes in three and four dimensions. These examples show how the same limiting prescription accommodates different horizon geometries and radial-order requirements.

Section~\ref{sec:mechanism} develops the general construction in Gaussian null coordinates. Section~\ref{sec:applications} applies this construction to different black hole geometries.
Section~\ref{sec:conclusions}  summarizes the results.
Appendix~\ref{app:pullback} collects the Gaussian null pullback
formulas through second radial order.

\section{General construction in Gaussian null coordinates}
\label{sec:mechanism}
We consider a smooth family of stationary black hole metrics in \(D\) spacetime dimensions, with nonextremal Killing horizons and a regular extremal limit. We work in a common Gaussian null chart in which the metric coefficients extend smoothly to extremality. We fix the normalization of the Killing generator of the outer horizon and denote the corresponding constant surface gravity by \(\kappa\). The parameter \(\varepsilon\) controls the radial expansion.
\subsection{non extremal near horizon expansion}
\label{sec:data}

In stationary Gaussian null coordinates $(v,\rho,x^A)$,\footnote{In the
general construction, $x=(x^1,\ldots,x^{D-2})$ denotes the coordinates on a
spatial horizon cross section.} with
$\cH=\{\rho=0\}$, $K=\partial_v$, and
$g(\partial_v,\partial_\rho)=1$, the metric takes the standard
form~\cite{FriedrichRaczWald,KunduriLucietti}
\begin{equation}
\begin{aligned}
\dd s^2={}&2\dd v\dd\rho+\rho q(\rho,x)\dd v^2\\
&+2\rho\omega_{A}(\rho,x)\dd v\dd x^A\\
&+\gamma_{AB}(\rho,x)\dd x^A\dd x^B .
\end{aligned}
\label{eq:gnc}
\end{equation}
Here $\partial_\rho$ is tangent to affinely parametrized transverse null
geodesics, and all coefficients are independent of $v$. With the convention
$K^\nu\nabla_\nu K^\mu=\kappa K^\mu$,
one has $q(0,x)=-2\kappa$. At fixed nonextremal black hole parameters, we write
\begin{equation}
\begin{aligned}
q&=-2\kappa+\rho F+O(\rho^2),\\
\omega_{A}&=\omega^{(0)}_{A}
 +\rho\omega^{(1)}_{A}+O(\rho^2),\\
\gamma_{AB}&=\gamma^{(0)}_{AB}
 +\rho\gamma^{(1)}_{AB}\\
&\quad+\frac{\rho^2}{2}\gamma^{(2)}_{AB}+O(\rho^3).
\end{aligned}
\label{eq:taylor}
\end{equation}
Thus $F=\tfrac12\partial_\rho^2g_{vv}|_{\cH}$,
$\omega^{(1)}_{A}=\tfrac12\partial_\rho^2g_{vA}|_{\cH}$,
and $\gamma^{(p)}_{AB}=\partial_\rho^p\gamma_{AB}|_{\cH}$.

We denote the metric after the radial scaling $\rho=\eps R$ by
$\dd s^2_\eps$. At fixed black hole parameters, its near horizon expansion
through second order is
\begin{equation}
\begin{aligned}
\dd s^2_{\eps}
 &=h+\eps k_1+\eps^2 k_2+O(\eps^3),\\
h&=\gamma^{(0)}_{AB}\dd x^A\dd x^B,\\
k_1&=2\dd v\dd R-2\kappa R\dd v^2\\
 &\quad+2R\omega^{(0)}_{A}\dd v\dd x^A\\
 &\quad+R\gamma^{(1)}_{AB}\dd x^A\dd x^B,\\
k_2&=R^2F\dd v^2
 +2R^2\omega^{(1)}_{A}\dd v\dd x^A\\
 &\quad+\frac{R^2}{2}\gamma^{(2)}_{AB}\dd x^A\dd x^B .
\end{aligned}
\label{eq:carroll}
\end{equation}
At leading order, only the transverse metric \(h\) contributes, while the longitudinal sector first appears at order \(\varepsilon\). The terms $h+\eps k_1$
give the string Carroll expansion~\cite{BagchiI,BagchiII}. Following ~\cite{ShindeMandlik}, we call its extension to higher radial orders
the non extremal near horizon expansion.

For a horizon regular seed metric, the Gaussian null coordinates are
constructed by the affine null geodesic map
\begin{equation}
\begin{aligned}
Y^\mu&=Y_H^\mu+\rho n^\mu+\frac{\rho^2}{2}\cA^\mu+O(\rho^3),\\
\cA^\mu&=-\Gamma^\mu{}_{\alpha\beta}n^\alpha n^\beta\big|_{\cH}.
\end{aligned}
\label{eq:map}
\end{equation}
Here $n$ is null, obeys $K\cdot n=1$, and is orthogonal to the
chosen horizon cross-section; $\cA^\mu$ is its coordinate acceleration.
The pullback formulas in Appendix~\ref{app:pullback} determine the
coefficients in Eq.~\eqref{eq:carroll} directly from the seed metric.

\subsection{near extremal scaling of the string Carroll expansion}
\label{sec:scaling}
Consider the first order truncation
$\dd s^2_{[1],\eps}=h+\eps k_1$ of the non extremal near horizon expansion
at fixed black hole parameters.\footnote{The label $[1]$ denotes the expansion of the metric through first order in $\eps$ which is the string Carroll expansion.} Simply taking $\kappa\to0$ in this truncation does
not, in general, recover the extremal near horizon geometry.

We instead consider a limit in which the approach to the outer
horizon and the approach to extremality occur simultaneously.
We implement this by scaling the surface gravity with the same
parameter $\eps$ introduced in the radial scaling $\rho=\eps R$,
such that
\begin{equation}
\frac{\kappa}{\eps}\longrightarrow\hat{\kappa}.
\label{eq:scaling}
\end{equation} Following ~\cite{ShindeMandlik}, we call this
the near extremal scaling. 
To retain a finite longitudinal sector, we also rescale Killing time as
\begin{equation}
v=\frac{V}{\eps},
\label{eq:time rescaling}
\end{equation}
and take $\eps\to0$ at fixed $(V,R,x^A)$ as the family approaches its extremal limit. The finite parameter $\hat{\kappa}$ is the
surface gravity in the rescaled time normalization.

The scaling of $\kappa$ keeps the linear temporal term finite under the radial scaling and time rescaling, while the transverse correction $\eps R\gamma^{(1)}_{AB}\dd x^A\dd x^B$ vanishes.
The limit of this first order truncation is\footnote{The metric subscript
$\hat{\kappa}$ labels the scaled surface gravity; a metric subscript $0$
denotes $\hat{\kappa}=0$. A subscript $0$ on horizon coefficients denotes
their extremal limiting values.}
\begin{equation}
\begin{aligned}
\dd s^2_{[1],\hat{\kappa}}={}&2\dd V\dd R-2\hat{\kappa} R\dd V^2\\
&+2R\omega^{(0)}_{0A}\dd V\dd x^A\\
&+\gamma^{(0)}_{0AB}\dd x^A\dd x^B.
\end{aligned}
\label{eq:first-order-limit}
\end{equation}
At $\hat{\kappa}=0$, let $\dd s^2_{\NH,0}$ denote the standard
extremal near horizon metric~\cite{KunduriLucietti} in the same Gaussian
null chart. Comparison gives
\begin{equation}
\dd s^2_{\NH,0}-\dd s^2_{[1],0}=R^2F_0(x)\dd V^2.
\label{eq:first-order-obstruction}
\end{equation}
Thus the string Carroll expansion fails to recover the extremal
near horizon metric when $F_0\not\equiv0$: it omits the contribution
of the second radial derivative of $g_{vv}$.

\subsection{second order contribution and limiting geometry}
\label{sec:orders}

Under the near extremal scaling~\eqref{eq:scaling} and the time
rescaling \eqref{eq:time rescaling}, the second order contribution becomes
\begin{equation}
\begin{aligned}
\eps^2 k_2={}&R^2F\dd V^2\\
&+2\eps R^2\omega^{(1)}_{A}\dd V\dd x^A\\
&+\frac{\eps^2R^2}{2}\gamma^{(2)}_{AB}\dd x^A\dd x^B\\
&\longrightarrow R^2F_0(x)\dd V^2.
\end{aligned}
\label{eq:promotion}
\end{equation}
The factor
$\dd v^2=\eps^{-2}\dd V^2$ promotes the longitudinal term to finite
order; the mixed and transverse terms vanish. Adding the surviving
term to Eq.~\eqref{eq:first-order-limit} yields the full limiting metric
\begin{equation}
\boxed{\begin{aligned}
\dd s^2_{\hat{\kappa}}={}&2\dd V\dd R
 +\bigl[R^2F_0(x)-2\hat{\kappa} R\bigr]\dd V^2\\
&+2R\omega^{(0)}_{0A}(x)\dd V\dd x^A\\
&+\gamma^{(0)}_{0AB}(x)\dd x^A\dd x^B .
\end{aligned}}
\label{eq:throat}
\end{equation}
The resulting metric is Lorentzian and nondegenerate. For $\hat{\kappa}>0$, it describes a near extremal throat with a
nondegenerate horizon at $R=0$, generated by $\partial_V$ and having surface gravity $\hat{\kappa}$. Setting $\hat{\kappa}=0$
in Eq.~\eqref{eq:throat} gives the extremal near horizon metric.

The radial coefficients remain finite as extremality is
approached. Under the radial scaling and time rescaling, the corresponding
terms scale as
\begin{equation}
\begin{aligned}
\rho^p\dd v^2&\sim\eps^{p-2},\\
\rho^p\dd v\dd x^A&\sim\eps^{p-1},\\
\rho^p\dd x^A\dd x^B&\sim\eps^p.
\end{aligned}
\label{eq:weights}
\end{equation}
The linear term in $g_{vv}$ remains finite because
$\kappa=O(\eps)$. Smoothness through extremality ensures that
the omitted radial remainder contributes $O(\eps)$ after
scaling and therefore vanishes at fixed $(V,R,x^A)$.
The surviving horizon coefficients approach their extremal
values, yielding Eq.~\eqref{eq:throat}.
Expansion through second radial order is therefore sufficient and necessary when
$F_0\not\equiv0$ in the chosen Gaussian null chart.
If $F_0\equiv0$, the string Carroll expansion suffices, as in
the neutral rotating BTZ example below.

In static coordinates that are singular
at the horizon, radial expansion coefficients can diverge as
extremality is approached, and arbitrarily high radial orders
can contribute, as demonstrated for RN in \cite{ShindeMandlik}.
\section{Applications to black hole geometries}
\label{sec:applications}

For each example, we evaluate $(\gamma^{(0)},\omega^{(0)},F)$
at fixed black hole parameters and insert their values at extremality into
Eq.~\eqref{eq:throat}. For the examples in four dimensions,
$r_\pm$ denote the outer and inner horizon radii.

\subsection{Reissner--Nordstr\"om black hole}
\label{sec:rn}

With \(\rho=r-r_+\), the RN solution~\cite{Reissner1916,Nordstrom1918}
takes the following ingoing Gaussian null form:
\begin{equation}
\begin{gathered}
\begin{aligned}
\dd s^2={}&2\dd v\dd\rho
 -\frac{\rho(\rho+\delta)}{(r_++\rho)^2}\dd v^2\\
&+(r_++\rho)^2\dd\Omega_2^2,
\end{aligned}\\
\delta=r_+-r_->0.
\end{gathered}
\label{eq:rn-seed}
\end{equation}
Direct comparison with \eqref{eq:taylor} gives
\begin{equation}
\begin{gathered}
\kappa=\frac{\delta}{2r_+^2},\qquad
F=-\frac{1}{r_+^2}+\frac{2\delta}{r_+^3},\\
\omega^{(0)}_{A}=0,\qquad
\gamma^{(0)}_{AB}\dd x^A\dd x^B=r_+^2\dd\Omega_2^2.
\end{gathered}
\label{eq:rn-data}
\end{equation}
For $r_\pm\to r_0>0$, the seed remains smooth in $\rho$ through
extremality. The near extremal scaling takes the form
$\delta/\eps\to2r_0^2\hat{\kappa}$. Since $F_0=-r_0^{-2}$,
Eq.~\eqref{eq:throat} yields
\begin{equation}
\begin{aligned}
\dd s^2_{\mathrm{RN},\hat{\kappa}}={}&2\dd V\dd R
 -\left(\frac{R^2}{r_0^2}+2\hat{\kappa} R\right)\dd V^2\\
&+r_0^2\dd\Omega_2^2.
\end{aligned}
\label{eq:rn-throat}
\end{equation}
This reproduces the scaled RN throat of~\cite{ShindeMandlik}
and is locally equivalent to the near extremal RN throat
described in~\cite{MaldacenaMichelsonStrominger}.
The two dimensional longitudinal sector has scalar curvature \(-2/r_0^2\); omitting
the second order term would instead give zero curvature. At $\hat{\kappa}=0$,
the local transformation $\dd V=\dd T+r_0^2\dd R/R^2$ identifies the
metric as extremal RN throat $\AdS_2(r_0)\times S^2(r_0)$~\cite{SpradlinStrominger}.

\subsection{Kerr black hole}
\label{sec:kerr}

Let $a_K$ be the rotation parameter of the Kerr solution~\cite{Kerr1963} and define
\begin{equation}
\begin{gathered}
\zeta=\cos\theta,\qquad u=1-\zeta^2,\\
\Delta_K=(r-r_+)(r-r_-),\qquad \Sigma_K=r^2+a_K^2\zeta^2,\\
\delta=r_+-r_-,\qquad P_K=r_+^2+a_K^2,\\
S_K=r_+^2+a_K^2\zeta^2,\qquad r_+r_-=a_K^2.
\end{gathered}
\label{eq:kerr-definitions}
\end{equation}
In advanced coordinates corotating with
$\Omega_H=a_K/P_K$, the non extremal seed metric is
\begin{equation}
\begin{aligned}
\dd s_K^2&=-\frac{\Delta_K}{\Sigma_K}U_K^2
 +2U_K\dd r+\frac{\Sigma_K}{u}\dd \zeta^2
 +\frac{u}{\Sigma_K}W_K^2,\\
U_K&=\frac{S_K}{P_K}\dd v-a_Ku\dd\phi,\\
W_K&=\frac{a_K(r^2-r_+^2)}{P_K}\dd v+(r^2+a_K^2)\dd\phi.
\end{aligned}
\label{eq:kerr-seed}
\end{equation}
The horizon generator is $\partial_v$. The seed radial coordinate is
not affine Gaussian null; the normalized transverse vector and its
radial coordinate acceleration in Eq.~\eqref{eq:map} are
\begin{equation}
\begin{aligned}
n&=\frac{a_K^2u}{2S_K}\partial_v
 +\frac{P_K}{S_K}\partial_r+\frac{a_K}{P_K}\partial_\phi,\\
\cA^r_K&=-\frac{a_K^2u(\delta S_K+4r_+P_K)}{2S_K^3}.
\end{aligned}
\label{eq:kerr-null}
\end{equation}
Here the components refer to the seed chart; after the pullback, $(v,\zeta,\phi)$
label the corresponding horizon coordinates. Applying
Appendix~\ref{app:pullback} gives
\begin{equation}
\begin{gathered}
\kappa=\frac{\delta}{2P_K},\\
\gamma^{(0)}_{AB}\dd x^A\dd x^B
 =\frac{S_K}{u}\dd \zeta^2+\frac{P_K^2u}{S_K}\dd\phi^2,\\
\begin{aligned}
\omega^{(0)}_{A}\dd x^A
 &=-\frac{2a_K^2\zeta}{S_K}\dd \zeta
 \\&\quad+\frac{a_Ku(\delta S_K+2r_+P_K)}{S_K^2}\dd\phi,
\end{aligned}\\
\begin{aligned}
F&=-\frac{1}{S_K}+\frac{2r_+\delta}{S_K^2}
 +\frac{4a_K^2r_+^2u}{S_K^3}\\
 &\quad+\frac{a_K^2u\delta(\delta S_K+4r_+P_K)}{4P_K^2S_K^2}.
\end{aligned}
\end{gathered}
\label{eq:kerr-data}
\end{equation}
For \(r_+\to r_0>0\), the Kerr metric and the transverse null geodesic construction remain smooth as \(\delta\to0\). The resulting Gaussian null expansion coefficients therefore have regular extremal limits.
We choose the orientation $a_K\to r_0$. The near extremal scaling
then gives $\delta/\eps\to4r_0^2\hat{\kappa}$. We define
$\Gamma_K=r_0^2(1+\cos^2\theta)$ and
$\Lambda_K=2\sin\theta/(1+\cos^2\theta)$. The extremal data are
\begin{equation}
\begin{aligned}
\gamma^{(0)}_{0AB}\dd x^A\dd x^B
 &=\Gamma_K(\dd\theta^2+\Lambda_K^2\dd\phi^2),\\
\omega^{(0)}_{0A}\dd x^A&=-\dd\ln\Gamma_K+\Lambda_K^2\dd\phi,\\
F_0&=\frac{\Lambda_K^2-1}{\Gamma_K}.
\end{aligned}
\label{eq:kerr-endpoint}
\end{equation}
Substituting these data into Eq.~\eqref{eq:throat}, we set \(R=\Gamma_K(\theta)y\) and define \(f_{\hat{\kappa}}(y)=y(y+2\hat{\kappa})\). On patches where $f_{\hat{\kappa}}\ne0$,
the coordinate changes
\begin{equation}
\begin{aligned}
\widetilde\phi&=\phi-\hat{\kappa} V,\\
\dd V&=\dd t+\frac{\dd y}{f_{\hat{\kappa}}(y)},\\
\dd\phi_s&=\dd\widetilde\phi+
 \frac{y+\hat{\kappa}}{f_{\hat{\kappa}}(y)}\dd y
\end{aligned}
\label{eq:kerr-static-map}
\end{equation}
put the metric in the familiar near NHEK form~\cite{BredbergHartmanSongStrominger,ChenChu}
\begin{equation}
\begin{aligned}
\dd s^2_{K,\hat{\kappa}}={}&\Gamma_K\left[-f_{\hat{\kappa}}(y)\dd t^2
 +\frac{\dd y^2}{f_{\hat{\kappa}}(y)}+\dd\theta^2\right]\\
&+\Gamma_K\Lambda_K^2[\dd\phi_s+(y+\hat{\kappa})\dd t]^2.
\end{aligned}
\label{eq:near-nhek}
\end{equation}
At $\hat{\kappa}=0$ this is the NHEK metric~\cite{BardeenHorowitz}.

\subsection{Kerr--Newman black hole}
\label{sec:kn}

For the Kerr--Newman solution~\cite{NewmanEtAl1965}, let $a$ denote the rotation parameter and retain
$\zeta=\cos\theta$, $u=1-\zeta^2$, and $\delta=r_+-r_-$.
We use mass and charge parameters $M$ and $Q$ with dimensions of length.
The horizon functions are
\begin{equation}
\begin{gathered}
\Delta=r^2-2Mr+a^2+Q^2=(r-r_+)(r-r_-),\\
r_+r_-=a^2+Q^2,\qquad \kappa=\frac{\delta}{2P},\\
P=r_+^2+a^2,\qquad S=r_+^2+a^2\zeta^2,\\
\Sigma=r^2+a^2\zeta^2.
\end{gathered}
\label{eq:kn-definitions}
\end{equation}
The advanced corotating seed has the form \eqref{eq:kerr-seed} with
$(a_K,P_K,S_K,\Sigma_K,\Delta_K)$ replaced by
$(a,P,S,\Sigma,\Delta)$. Direct evaluation of the same pullback gives
\eqref{eq:kerr-null} and \eqref{eq:kerr-data} with those replacements.
The common expressions follow from $\Delta|_H=0$,
$\Delta'|_H=\delta$, and $\Delta''|_H=2$; charge enters through the
horizon relation in Eq.~\eqref{eq:kn-definitions}.

Let $r_\pm\to r_0>0$, $a\to a_0$, and $Q\to Q_0$. The non extremal
horizon identity implies $r_0^2=a_0^2+Q_0^2$. We then define
\begin{equation}
\begin{gathered}
P_0=r_0^2+a_0^2,\qquad \Gamma=r_0^2+a_0^2\cos^2\theta,\\
\Lambda=\frac{P_0\sin\theta}{\Gamma},\qquad
k=\frac{2a_0r_0}{P_0}.
\end{gathered}
\label{eq:kn-functions}
\end{equation}
The functions $\Lambda$ and $\Lambda_K$ are both dimensionless.
The near extremal scaling gives
$\delta/\eps\to2P_0\hat{\kappa}$, and the required extremal data are
\begin{equation}
\begin{gathered}
\gamma^{(0)}_{0AB}\dd x^A\dd x^B=\Gamma(\dd\theta^2+\Lambda^2\dd\phi^2),\\
\omega^{(0)}_{0A}\dd x^A=-\dd\ln\Gamma+k\Lambda^2\dd\phi,\\
F_0=\frac{k^2\Lambda^2-1}{\Gamma}.
\end{gathered}
\label{eq:kn-endpoint}
\end{equation}
Substituting these data into \eqref{eq:throat}, we set \(R=\Gamma y\). The time transformation in \eqref{eq:kerr-static-map}, together with
\[
\begin{aligned}
\widetilde{\phi}&=\phi-k\hat{\kappa}V,\\
\dd\phi_s&=\dd\widetilde{\phi}
+k\,\frac{y+\hat{\kappa}}{f_{\hat{\kappa}}(y)}\,\dd y,
\end{aligned}
\]
puts the metric in the near NHEKN form \cite{ChenChu},
\begin{equation}
\begin{aligned}
\dd s^2_{\mathrm{KN},\hat{\kappa}}
={}&\Gamma\left[
-f_{\hat{\kappa}}(y)\dd t^2
+\frac{\dd y^2}{f_{\hat{\kappa}}(y)}
+\dd\theta^2
\right]\\
&+\Gamma\Lambda^2
\left[\dd\phi_s+k(y+\hat{\kappa})\dd t\right]^2.
\end{aligned}
\label{eq:near-nhekn}
\end{equation}
At $\hat{\kappa}=0$ this is NHEKN~\cite{BardeenHorowitz}. The Kerr limit $Q_0=0$, $a_0=r_0$ gives $k=1$,
$\Gamma=\Gamma_K$, and $\Lambda=\Lambda_K$,
recovering the Kerr throat in Eq.~\eqref{eq:near-nhek}. The nonrotating limit $a_0=0$ gives $k=0$, $\Gamma=r_0^2$, and
$\Lambda=\sin\theta$, recovering the RN throat in
Eq.~\eqref{eq:rn-throat} with $R=r_0^2y$.

\subsection{Charged rotating BTZ black hole}
\label{sec:charged-btz}

We use the charged rotating BTZ solution~\cite{ClementChargedBTZ,MartinezTeitelboimZanelli,MaedaPodolsky}
with $\AdS_3$ radius $\ell$ and Maxwell normalization
$-\tfrac14\mathcal F_{\mu\nu}\mathcal F^{\mu\nu}$. We set
\begin{equation}
\begin{gathered}
s=1-\beta^2>0,\qquad \alpha=8\pi G_3Q^2,\\
f(r)=\frac{r^2}{\ell^2}-m-\alpha\ln\frac{r}{r_\star},\\
C^2(r)=\frac{r^2-\beta^2\ell^2 f(r)}{s},\\
N^\varphi(r)=\frac{\beta\ell}{sC^2(r)}
 \left[f(r)-\frac{r^2}{\ell^2}\right].
\end{gathered}
\label{eq:btz-functions}
\end{equation}
Here $0\leq\beta<1$, $(G_3,\ell,r_\star)$ are fixed, and $m,Q$
are seed parameters. In the region $r>0$, $C^2>0$, the metric is
\begin{equation}
\begin{gathered}
\begin{aligned}
\dd s^2={}&-\frac{r^2f}{C^2}\dd t^2+\frac{\dd r^2}{f}\\
&+C^2(\dd\widehat\varphi+N^\varphi\dd t)^2,
\end{aligned}\\
\widehat\varphi\sim\widehat\varphi+2\pi .
\end{gathered}
\label{eq:btz-seed}
\end{equation}
Let $r_h$ be the outer root of $f$, with $f'_h>0$. Then
$C_h=r_h/\sqrt{s}$ and $\Omega_H=\beta/\ell$. Following the
horizon regular coordinate construction in ~\cite{MaedaPodolsky}
and passing to a corotating angle, the transformations
\begin{equation}
\begin{aligned}
\dd v&=\dd t+\frac{C}{rf}\dd r,\\
\dd\widetilde\varphi&=\dd\widehat\varphi
 -\frac{CN^\varphi}{rf}\dd r,\\
\varphi&=\widetilde\varphi-\Omega_Hv,\qquad
\rho=\int_{r_h}^{r}\frac{z}{C(z)}\dd z
\end{aligned}
\label{eq:btz-map}
\end{equation}
give the exact Gaussian null metric
\begin{equation}
\begin{aligned}
\dd s^2={}&2\dd v\dd\rho-sf(r(\rho))\dd v^2\\
&+2\beta\ell f(r(\rho))\dd v\dd\varphi
 +C^2(r(\rho))\dd\varphi^2.
\end{aligned}
\label{eq:btz-gnc}
\end{equation}
The affine radial map is regular at the horizon, where
$\dd\rho/\dd r=\sqrt{s}$. Using
$\partial_\rho=(C/r)\partial_r$ and
$\eta=\alpha\ell^2/(2r_h^2)$, the nonextremal horizon data are
\begin{equation}
\begin{aligned}
\kappa&=\frac{\sqrt{s}\,r_h}{\ell^2}(1-\eta),\\
F&=-\frac{1+\eta-\beta^2(1-\eta)^2}{\ell^2},\\
\omega^{(0)}_{\varphi}
 &=\frac{2\beta r_h}{\ell\sqrt{s}}(1-\eta)
 =\frac{2\beta\ell}{s}\kappa,\\
\gamma^{(0)}_{\varphi\varphi}&=\frac{r_h^2}{s}.
\end{aligned}
\label{eq:btz-data}
\end{equation}
For a charged outer horizon, $0<\eta<1$. Along a family satisfying the
near extremal scaling, with
$r_h\to r_0>0$, $Q\to Q_0\ne0$, $\beta\to\beta_0$, and
$s\to s_0=1-\beta_0^2>0$, we have
\begin{equation}
\begin{gathered}
\frac{1-\eta}{\eps}\longrightarrow
\frac{\ell^2\hat{\kappa}}{\sqrt{s_0}\,r_0},\\
\alpha_0=\frac{2r_0^2}{\ell^2}.
\end{gathered}
\label{eq:btz-endpoint-relations}
\end{equation}
Thus $\eta\to1$. With $C_0=r_0/\sqrt{s_0}$, the extremal data are
\begin{equation}
\begin{gathered}
F_0=-\frac{2}{\ell^2},\qquad \omega^{(0)}_{0\varphi}=0,\\
\gamma^{(0)}_{0\varphi\varphi}=C_0^2.
\end{gathered}
\label{eq:btz-endpoint}
\end{equation}
Since $r_0,s_0>0$, the seed functions and affine inverse remain smooth
through extremality. Equation~\eqref{eq:throat} yields
\begin{equation}
\begin{aligned}
\dd s^2_{\mathrm{chBTZ},\hat{\kappa}}={}&2\dd V\dd R
 -\left(2\hat{\kappa} R+\frac{2R^2}{\ell^2}\right)\dd V^2\\
&+C_0^2\dd\varphi^2.
\end{aligned}
\label{eq:charged-btz-throat}
\end{equation}
At \(\hat{\kappa}=0\), this is the charged BTZ near horizon metric \(\mathrm{AdS}_2(\ell/\sqrt{2})\times S^1(C_0)\)~\cite{ClementFabbriThroats,KunduriLucietti,Cadoni:2008mw}. For all \(\hat{\kappa}\geq0\), the rotational fibration is absent \cite{MatyjasekZaslavskii}, and \(\beta_0\) enters only through the circle radius \(C_0\). Setting \(\beta_0=0\) gives \(C_0=r_0\) and recovers the nonrotating charged BTZ throat.
\subsubsection{Neutral rotating BTZ black hole}
\label{sec:neutral-btz}

For the neutral rotating BTZ family~\cite{BanadosTeitelboimZanelli,BanadosHenneauxTeitelboimZanelli},
set $Q=0$ before taking the extremal limit and
introduce the horizon circumference radii
\begin{equation}
\begin{gathered}
C_+=\frac{r_h}{\sqrt{s}},\qquad C_-=\beta C_+,\\
\kappa=\frac{C_+^2-C_-^2}{\ell^2C_+}.
\end{gathered}
\label{eq:neutral-radii}
\end{equation}
Now $C^2=r^2+C_-^2$, so $\rho=C-C_+$. Equation \eqref{eq:btz-gnc}
becomes 
\begin{equation}
\begin{aligned}
\dd s^2={}&2\dd v\dd\rho
 -\frac{\kappa}{C_+}\rho(2C_++\rho)\dd v^2\\
&+\frac{2C_-}{\ell C_+}\rho(2C_++\rho)\dd v\dd\varphi\\
&+(C_++\rho)^2\dd\varphi^2.
\end{aligned}
\label{eq:neutral-gnc}
\end{equation}

Hence $F=-\kappa/C_+$,
$\omega^{(0)}_{\varphi}=2C_-/\ell$, and
$\gamma^{(0)}_{\varphi\varphi}=C_+^2$.
In the variables $C_\pm$,
Eq.~\eqref{eq:neutral-gnc} is a smooth Gaussian null family to which
the same near extremal scaling applies.
For $C_\pm\to C_0>0$ with $\kappa/\eps\to\hat{\kappa}$, the extremal
data and the resulting metric are
\begin{equation}
\begin{gathered}
F_0=0,\qquad \omega^{(0)}_{0\varphi}=\frac{2C_0}{\ell},\qquad
\gamma^{(0)}_{0\varphi\varphi}=C_0^2,\\
\begin{aligned}
\dd s^2_{\mathrm{nBTZ},\hat{\kappa}}={}&2\dd V\dd R-2\hat{\kappa} R\dd V^2\\
&+\frac{4C_0R}{\ell}\dd V\dd\varphi+C_0^2\dd\varphi^2.
\end{aligned}
\end{gathered}
\label{eq:neutral-throat}
\end{equation}
Since \(F_0=0\), the second order contribution vanishes in the limit, and the string Carroll expansion already gives the complete limiting metric. At \(\hat{\kappa}=0\), \eqref{eq:neutral-throat} reproduces the extremal rotating BTZ near horizon geometry, a circle fibration over \(\mathrm{AdS}_2(\ell/2)\), locally \(\mathrm{AdS}_3(\ell)\) \cite{CoussaertHenneauxSelfDual,StromingerAdS2,BalasubramanianChiralBTZ}.

In this limit, \(\beta\to1\), \(s\to0\), and \(r_h\to0\), while \(C_0\) remains finite. This lies outside the charged extremal regime \(r_0,s_0>0\) \cite{MartinezTeitelboimZanelli,MaedaPodolsky}. Consequently, the neutral throat cannot be obtained by simply setting \(Q_0=0\) in Eq.~\eqref{eq:charged-btz-throat}.

\section{Conclusions}
\label{sec:conclusions}
In this paper we have given a general prescription for recovering extremal near horizon geometries from the non extremal near horizon expansion. The charged and rotating examples extend the earlier RN construction~\cite{ShindeMandlik} beyond spherical symmetry, with a common pullback procedure supplying the coefficients required by the limit.

The significance of the result lies in the reorganization of the expansion near extremality. Terms that are subleading at fixed black hole parameters can contribute at leading order under the near extremal scaling and the time rescaling. Consequently, the string Carroll expansion need not retain all the information required for the throat. The second order analysis identifies the contribution that restores the limiting geometry and the condition under which it must be included.

A natural direction for future work is to explore the implications of this construction for probe dynamics. This would extend the geometric connection established here to the behavior of particles, fields, and strings.

\acknowledgments
We thank the people of India for their continuous support for science. We thank IIT(ISM) Dhanbad and the Department of Physics for providing us with the academic infrastructure necessary for our research work.

\appendix
\section{Gaussian null pullback through second radial order}
\label{app:pullback}

We work at fixed black hole parameters. In stationary,
corotating seed coordinates, let $Y_H^\mu=(v,r_h,x^A)$ and
$K=\partial_v$. For any seed metric component $\Psi=g_{\mu\nu}$,
the map~\eqref{eq:map} gives
\begin{equation}
\begin{aligned}
\Psi(Y)={}&[\Psi]_H+\rho[n^\alpha\partial_\alpha \Psi]_H\\
&+\frac{\rho^2}{2}\bigl[\cA^\alpha\partial_\alpha \Psi
 +n^\alpha n^\beta\partial_\alpha\partial_\beta \Psi\bigr]_H\\
&+O(\rho^3).
\end{aligned}
\label{eq:seed-taylor}
\end{equation}
The geodesic map respects the stationary Killing flow, so
$\partial_vY^\mu=\delta_v{}^\mu$, while
the transverse Jacobian is
\begin{equation}
\begin{aligned}
\partial_A Y^\mu={}&\delta_A{}^\mu+\rho\partial_A n^\mu\\
&+\frac{\rho^2}{2}\partial_A\cA^\mu+O(\rho^3).
\end{aligned}
\label{eq:jacobian}
\end{equation}
All quantities on the right-hand sides below are seed components,
evaluated on the horizon after differentiation. The coefficients
entering the limiting metric are
\begin{align}
\gamma^{(0)}_{AB}&=[g_{AB}]_H,
\label{eq:pull-gamma0}\\
\omega^{(0)}_A&=
 \bigl[n^\alpha\partial_\alpha g_{vA}
 +g_{v\mu}\partial_A n^\mu\bigr]_H,
\label{eq:pull-omega0}\\
F&=\frac12\bigl[\cA^\alpha\partial_\alpha g_{vv}
 +n^\alpha n^\beta\partial_\alpha\partial_\beta g_{vv}\bigr]_H.
\label{eq:pull-f}
\end{align}
The remaining coefficients in \eqref{eq:carroll} follow from the same
pullback:
\begin{equation}
\begin{aligned}
\gamma^{(1)}_{AB}
 &=\bigl[n^\alpha\partial_\alpha g_{AB}
 +g_{\mu B}\partial_A n^\mu\\
 &\qquad+g_{A\nu}\partial_B n^\nu\bigr]_H,
\end{aligned}
\label{eq:pull-gamma1}
\end{equation}
\begin{equation}
\begin{aligned}
\omega^{(1)}_A
 &=\frac12\bigl[\cA^\alpha\partial_\alpha g_{vA}
 +n^\alpha n^\beta\partial_\alpha\partial_\beta g_{vA}\bigr]_H\\
 &\quad+\bigl[(n^\alpha\partial_\alpha g_{v\mu})\partial_A n^\mu\bigr]_H\\
 &\quad+\frac12\bigl[g_{v\mu}\partial_A\cA^\mu\bigr]_H,
\end{aligned}
\label{eq:pull-omega1}
\end{equation}
\begin{equation}
\begin{aligned}
\gamma^{(2)}_{AB}
 &=\bigl[\cA^\alpha\partial_\alpha g_{AB}
 +n^\alpha n^\beta\partial_\alpha\partial_\beta g_{AB}\bigr]_H\\
 &\quad+2\bigl[(n^\alpha\partial_\alpha g_{\mu B})\partial_A n^\mu\\
 &\qquad\quad+(n^\alpha\partial_\alpha g_{A\nu})\partial_B n^\nu\bigr]_H\\
 &\quad+2\bigl[g_{\mu\nu}\partial_A n^\mu\partial_B n^\nu\bigr]_H\\
 &\quad+\bigl[g_{\mu B}\partial_A\cA^\mu
 +g_{A\nu}\partial_B\cA^\nu\bigr]_H.
\end{aligned}
\label{eq:pull-gamma2}
\end{equation}
The second order geodesic map determines \(g_{vv}\), \(g_{vA}\), and \(g_{AB}\) through second radial order. The Gaussian null construction fixes the remaining components of the pulled back metric exactly to \(g_{v\rho}=1\) and \(g_{\rho\rho}=g_{\rho A}=0\).

\renewcommand{\bibsection}{\section*{\refname}}
\interlinepenalty=10000 
\bibliography{references}
\end{document}